\documentclass[journal,transmag]{IEEEtran}
\usepackage{amssymb}
\usepackage{bm}
\usepackage{booktabs}
\usepackage{threeparttable}
\usepackage{algorithmic}
\usepackage{multirow}
\usepackage{graphicx}
\usepackage{algorithm}
\usepackage{pifont}
\usepackage{amsmath}
\usepackage{xcolor,enumerate}
\usepackage{makecell}
\usepackage{multirow}
\usepackage{multicol}
\usepackage{enumitem}
\usepackage{wasysym}
\usepackage{framed}
\usepackage{pgfgantt,rotating}
\usepackage{subfloat}
\usepackage{multirow}
\usepackage{blindtext}
\usepackage[noadjust]{cite}
\usepackage{url}
\usepackage{algorithm,algorithmic}
\ifCLASSINFOpdf
\else
\fi

\begin{document}
%
\title{Design Ontology Supporting Model-based Systems-engineering Formalisms}


\author{\IEEEauthorblockN{Jinzhi Lu\IEEEauthorrefmark{1},~\IEEEmembership{CSEP},
Junda Ma\IEEEauthorrefmark{2},
Xiaochen Zheng\IEEEauthorrefmark{1},
Guoxin Wang\IEEEauthorrefmark{2}, and
Dimitris Kiritsis\IEEEauthorrefmark{1}}
\IEEEauthorblockA{\IEEEauthorrefmark{1} SCI STI DK,
Ecole Polytechnique Fédérale de Lausanne, Lausanne, 1015, Switzerland}
\IEEEauthorblockA{\IEEEauthorrefmark{2} Beijing Institute of Technology, Beijing, 100081, China} 

 \thanks{Corresponding author: Guoxin Wang is an associate Professor, Beijing Institute of Technology, Beijing, China (e-mail: wangguoxin@bit.edu.cn)}}%

\markboth{Journal of \LaTeX\ Class Files,~Vol.~14, No.~8, August~2015}%
{Shell \MakeLowercase{\textit{et al.}}: Bare Demo of IEEEtran.cls for IEEE Transactions on Magnetics Journals}

\IEEEtitleabstractindextext{%
\begin{abstract}
Model-based systems engineering (MBSE) provides an important capability for managing the complexities of system development. MBSE empowers the formalisms of system architectures for supporting model-based requirement elicitation, specification, design, development, testing, fielding, etc. However, the modeling languages and techniques are quite heterogeneous, even within the same enterprise system, which creates difficulties for data interoperability. The discrepancies among data structures and language syntaxes make information exchange among MBSE models even more difficult, resulting in considerable information deviations when connecting data flows across the enterprise. For this reason, this paper presents an ontology based upon \textit{graphs}, \textit{objects}, \textit{points}, \textit{properties}, \textit{roles}, and \textit{relationships} with \textit{entensions} (GOPPRRE), providing meta models that support the various lifecycle stages of MBSE formalisms. In particular, knowledge-graph models are developed to support unified model representations to further implement ontological data integration based on GOPPRRE throughout the entire lifecycle. The applicability of the MBSE formalism is verified using quantitative and qualitative approaches. Moreover, the GOPPRRE ontologies are generated from the MBSE language formalisms in a domain-specific modeling tool, \textit{MetaGraph} in order to evaluate its availiablity. The results demonstrate that the proposed ontology supports both formal structures and the descriptive logic of the systems engineering lifecycle.

\end{abstract}

}

\maketitle

\IEEEdisplaynontitleabstractindextext

\IEEEpeerreviewmaketitle

\section{Introduction}

\IEEEPARstart{T}he increasing complexity of technological innovations and their interoperability requirements within systems of systems, systems, subsystems and components, have led to an over-complexity of architectures and data structures, which, in turn, has led to enormous research and development costs. The model-based systems engineering (MBSE) has been widely used to counter this trend by formalizing end-to-end systems engineering perspectives through models. Each interface between lifecycle phases poses communication challenges brought about by this increasing complexity \cite{9062568}. Much of the complexity is the result of individual stakeholder interests; they may have different concerns about systems and artifacts of interest, and they may, in turn, demand unique informational and data-standard feedback. These results can often be seen within the architectural models themselves, as discrepancies among such models create a system-integration nightmare, resulting in barriers to communications, understandability, and, more importantly, operations. Apart from stakeholder nuances, the integration of model views is also challenged by different domain-specific knowledge base and systems-engineering taxonomies.

Across the entire lifecycle, enterprise data integration is the ultimate goal for a fully implemented MBSE. However, at the working levels, it is common for domain engineers to formalize their various domain problems using stove-piped domain-specific modeling languages and specifications. These various representations are difficult to piece together during collaborative development, and the results often lead to misinterpreted or inaccurate reporting. Therefore, there is a critical need to standardize information representations and data structures so that a complete model flow can be wielded across the lifecycle while meeting all stakeholder and engineering requirements.

System development is an iterative process that relies on a unified and authoritative data architecture built upon collaboration. Owing to advances in artificial intelligence (AI) and machine-learning (ML) techniques, the concept of MBSE is undergoing a digital transformation that will ultimately lead to advanced facilitation to complex system development \cite{McDermott2020}. Semantics web-data exchange is the basis for much current data and information integration via AI reasoning. Thus, it is critical that participants of the MBSE lifecycle work to ensure the completeness and consistency of the data that fuel decision-making and engineering task implementation. Based upon on an AI-driven data exchange, the knowledge management of the future aims to provide the required information to stakeholders whenever (or even before) they need it \cite{Hao2019}. 

This paper focuses on a unified MBSE ontology based on a meta-meta model built upon six key concepts with extensions: \textbf{G}raph; \textbf{O}bject; \textbf{P}oint; \textbf{P}roperty; \textbf{R}ole; and \textbf{R}elationship (GOPPRRE). This ontology presents a formalization opportunity for MBSE modeling via a unified syntax and data structure to support systems-engineering information exchange via the integration of AI and ML. The main contributions of this defined ontology are as follows:
\begin{itemize}
\item It supports integrated architectural representation across the lifecycle.
\item It promotes a MSBE tool built upon data interoperability and consistency.
\item It provides potential solutions for developing AI/ML MBSE roadmaps.
\end{itemize}
In order to promote the scalability of the proposed ontology, it will be discussed and applied in the Industrial Ontologies Foundry Systems Engineering Working Group \footnote{https://www.industrialontologies.org/}.

The rest of the paper is organized as follows. We discuss related works and the proposed research methodology in Section \uppercase\expandafter{\romannumeral2}. In Section \uppercase\expandafter{\romannumeral3}, the designed ontology is analyzed. A case study is presented in Section \uppercase\expandafter{\romannumeral4} for the evaluation of our ontology using quantitative and qualitative approaches. Finally, we present our conclusions in Section \uppercase\expandafter{\romannumeral5}. 

\section{Research Design}

\subsection{Literature review}

Some researchers have provided ontology-based approaches to facilitating design automation for complex systems (\cite{Ming2018}, \cite{Wang2019}). MBSE supports complex systems engineering and development efforts \cite{Dickerson2013} by formalizing development processes, system architectures, and operational interrelationships. There are currently several such modeling languages in use (e.g., Systems Modeling Language (SysML) \cite{Bassi2011}, Object Process Methodology \cite{Mordecai2018}, and Business Process Modeling Notation (BPMN) \cite{Ravikumar2018}), which provide modeling tools that can be used to describe real-world processes using graphic views. Recently, researchers have proposed an Object Management Group standard for model-driven engineering, comprising a four-layered architecture. The four layers are labeled M0--M3 and provide the modeling framework needed to support MBSE. The M0--M3 layers are described thoroughly in the ``Ontology Design for MBSE Formalism’’ section of this paper. It applies the GOPPRR formalization of specific system views with new extensions \cite{Kelly1996}. Notably, several generic modeling environments also provide meta-modeling languages that can support complex system development based on unified modeling-language (UML) notation and object constraint language  \cite{Nuzzo2015}. Advocates of these methods continue to seek a singular adaptive language that can be used to describe all system architecture views.

Yang \textit{et al.} provided a unified ontology to describe a systems-engineering body of knowledge for the International Council on Systems Engineering (\cite{Beihoff2010}, \cite{Yang2020}).  Charlotte \textit{et al.} proposed a formal method of safety analyses for systems engineering \cite{Seidner2008}. But these research were not involved with MBSE. Lu \textit{et al.} developed an ontology to support automated co-simulation using an MBSE tool-chain. The ontology was used to implement MBSE models for integrated verification \cite{Lu2020ieeeesss}. Most of the above ontological approaches, however, focused on domain-specific problems instead of modeling languages and data interoperability across the entire MBSE lifecycle.

MBSE was the basis for constructing a digital replication technologies and supporting virtual verification concepts (\cite{Schluse2018}, \cite{Wang2019}). It is further expected to provide potential solutions for combining systems engineering approaches and AI technologies. Some researchers provide an ontology-based approach facilitating design automation for complex systems (\cite{Ming2018}, \cite{Wang2019}). Hao \textit{et al.} proposed an ontology-based method to support knowledge management \cite{Hao2014}. Ontology contributes to semantics descriptions and models that not only support decision-making regarding system development, it also supports real-time operations via a universal 
system description and information transfer \cite{JINZHIct}.

Currently, ontological methods are widely used to support lower-level tool and data interoperability and consistency issues. For example, an extensible XML Metadata Interchange (XMI) is used to support data exchange between SysMLs and multiple other tools \cite{Papakonstantinou2013}. Additionally, MBSE ontologies have been developed to formalize domain-specific concepts and their interrelationships using different languages (\cite{Wang2019}, \cite{Walter2014}). In this paper, Web Ontology Language (OWL) is used to design a complete MBSE ontology based on a GOPPRRE approach that can support information exchange across the MBSE enterprise.

\subsection{Summary}

Several modeling languages have been used to formalize the different views and approaches found in the systems engineering lifecycle. Many challenges arise, however, when these different languages are adopted for different enterprises. 

\begin{itemize}
\item Generic modeling languages have difficulty supporting the complete formalism of a specific domain; they do not support multiple system views in a unified way. 
\item Different language models pose integration challenges across different development phases. 
\item An ontology that supports MBSE formalisms will be the basis of the combination of systems-engineering processes and AI tools for enterprise knowledge management and decision-makings.
\end{itemize}

\subsection{Case study}

A case study was conducted to evaluate the designed MBSE ontology. Quantitative and qualitative approaches were separately applied \cite{Lu2020ieeeesss}, and two key measurements were considered: 

\textit{\textbf{1. The ontological completeness of the concrete syntax of MBSE formalisms:}} 
\begin{itemize}
\item 
In the qualitative evaluation, SPARQL \cite{Patti2016}, a query language, was used to evaluate whether the ontology could completely represent the information generated from the MBSE models. To support this measurement, several metrics were defined:
\begin{itemize}
\item \textit{Graph-include-objects (relationship)} refers to a situation, wherein one model includes all information related to its components or connections.
\item \textit{Object (relationship)-include-points (roles)} refers to a situation wherein one model component or connection includes all information related to its points or connection arrows.
\item \textit{Object (graph and relationship)-include-properties} refers to a situation wherein one model (model component and connection) includes all information related to its attitudes. 
\end{itemize}

\item Using the quantitative approach, a domain-specific modeling tool, \textit{MetaGraph}, was developed to support the required ontology generation \cite{LU202020202020}. The numbers of key elements in the modeling languages and specifications for which the ontology was formalized were analyzed to evaluate its completeness.
\end{itemize}

\textit{\textbf{2. Ontology logic related to the abstract syntax of MBSE formalisms:}}
\begin{itemize}
\item In the qualitative evaluation, SQWRL \cite{OConnor2009}, a query language, was used to evaluate its description logic, by querying OWL as its semantic web-rule language to design the rules needed to assign the \textit{subject}, \textit{predicate}, and \textit{object} by their defined \textit{predicates} \cite{horrocks2004swrl}. It was adopted to evaluate whether the ontology could capture the information needed to define the abstract syntax of the MSBE models. Two metrics were thus considered:
\begin{itemize}
\item \textit{Relationship definitions in the MBSE model}: they present the connections (logic flows) with two ends in the MBSE models. The connections between model components or points define the basic logic for constructing one MBSE model. 
\item \textit{Direction of relationship}: it presents the start of each connection, which decides how the connection is linked to its two sides.
\end{itemize}
\item  Using the quantitative approach, \textit{MetaGraph} was used to support ontology generation, wherein the numbers of key graphs could support connection rules of different modeling languages and specifications. They were identified to evaluate the logic supported by the designed ontology.
\end{itemize}

\section{Ontology Design for MBSE Formalism}

\begin{figure}[!t]
\centering
\includegraphics[width=3.4in]{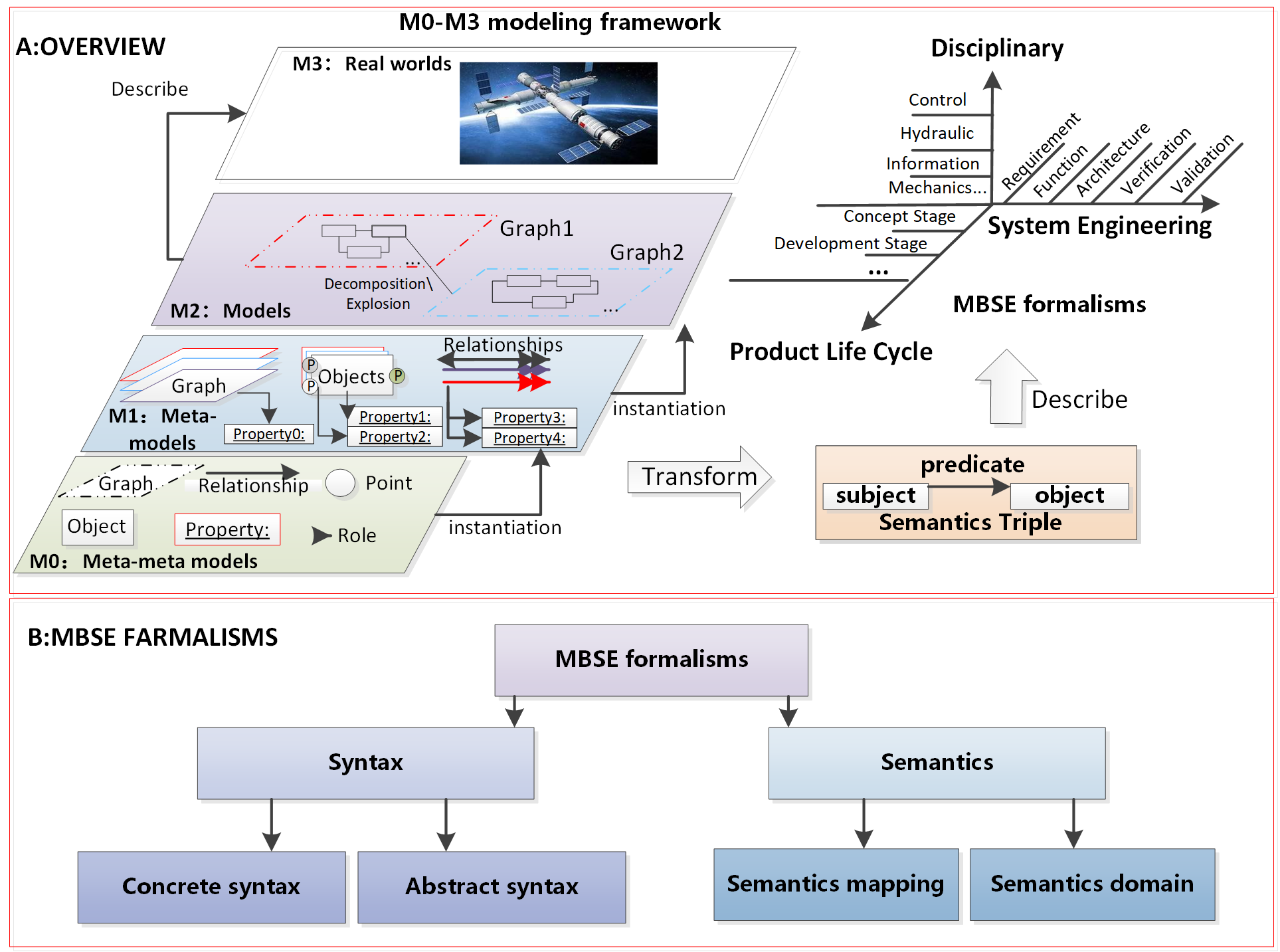}
\caption{Overview of the ontology and MBSE formalisms}
\label{F:Overviews}
\end{figure}

\subsection{Overview}
The overall workflow of the proposed approach is shown in Fig. \ref{F:Overviews}-A. The M0--M3 modeling framework is proposed to develop the MBSE ontology, including:
\begin{itemize}
\item M0: Meta-meta models that refer to basic elements of the constructed model compositions and their interconnections. We adopt GOPPRR meta-meta models and their extensions to support meta-model development.
\item M1: Meta-models refer to the model compositions and connections needed to develop models.
\item M2: MBSE models represent real-world systems.
\item M3: Real-world artifacts are considered, including complex systems and their development processes.
\end{itemize}
The developed ontology is transformed into semantics triples (i.e., subject, object, and predicate) to formalize systems-engineering models from three dimensions: disciplinary, system lifecycle, and system artifacts \cite{Lu2018aaaa}. The disciplinary dimension includes several domains, such as control engineering, mechanical engineering, etc. Systems engineering refers to requirements, functions, architectures, etc., and the product lifecycle includes different phases of the complex system lifecycle.

As shown in Fig. \ref{F:Overviews}-B, the MBSE formalisms include syntax and semantics \cite{Mierlo2017}. Syntax refers to the representations of the MBSE formalisms, and semantics refers to the MBSE model meanings. The details are explained as follows:
\begin{itemize}
\item Abstract syntax refers to the compositions of MBSE models and their defined rules for connecting with each composition. It is realized using the core GOPPRRE concepts for the MSBE formalisms (introduced in Section III.B)
\item Concrete syntax refers to the visual representations of the MBSE model compositions. It is represented in the knowledge-graph model as the annotation property, which is introduced in Section III.C. 
\item Semantics domain refers to the target of the semantic mapping, which implies the meanings of the MBSE models. It includes the three dimensions shown in Fig. \ref{F:Overviews}-A. The formalisms are used to describe system-engineering concepts, product-lifecycle processes, and disciplinary knowledge needed to support information exchange during system development.
\item Semantics mapping refers to the dependencies among MBSE models and their meanings according to the three dimensions. 
\end{itemize}

\begin{figure}[!t]
\centering
\includegraphics[width=3.4in]{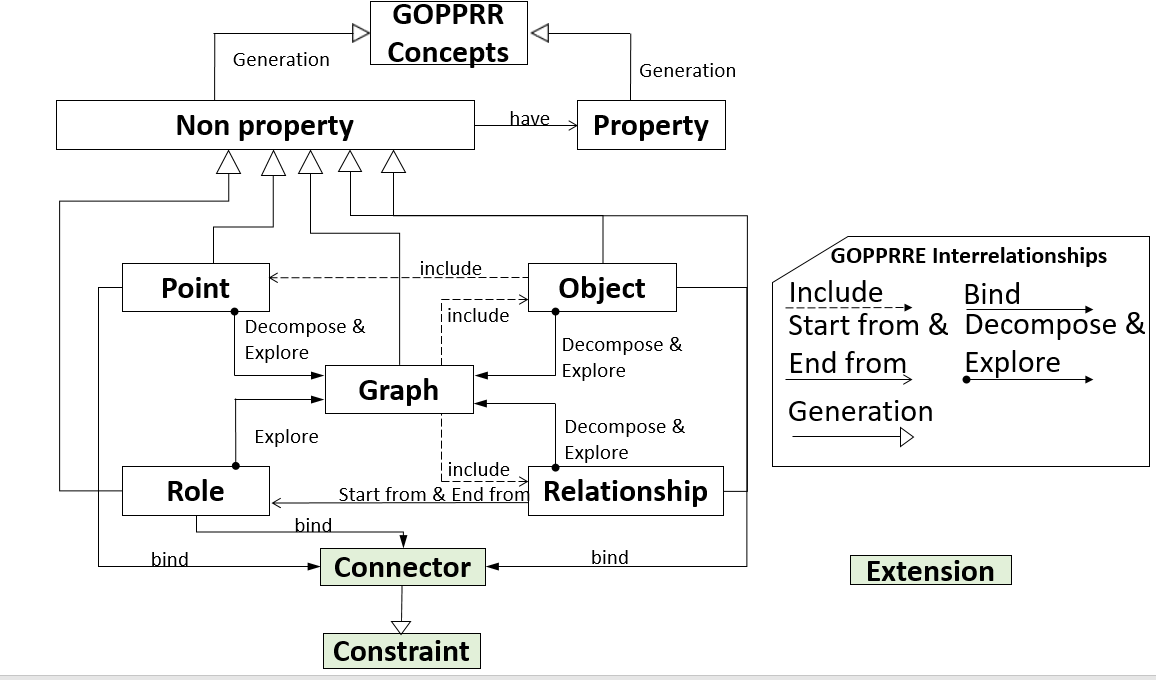}
\caption{Meta-models of GOPPRRE}
\label{F:GOPPRRE}
\end{figure}

\subsection{GOPPRRE Concepts for MBSE Formalisms}

The \textit{GOPPRRE} approach uses the M0--M3 modeling framework, as inspired by the GOPPRR meta-meta models and their extensions, to construct the MBSE model syntax and semantics. We added one new concept, \textit{constraint}, to the approach in order to define constraints of the abstract syntax. In Fig. \ref{F:GOPPRRE}, the details of the GOPPRRE concepts are introduced: 
\begin{itemize}
\item \textit{\textbf{G}raph} is an entity collection of \textit{Object}, \textit{Relationship}, and \textit{Role}, represented in one layout (e.g., a UML class diagram). The graph is either a visual diagram or another that was decomposed (explored) by one \textit{Object}.
\item \textit{\textbf{O}bject} is an entity that constructs a \textit{Graph}.
\item \textit{\textbf{P}oint} is one attached port in an \textit{Object}.
\item \textit{\textbf{R}elationship} refers to one connection between the \textit{Points} and/or \textit{Objects}.
\item \textit{\textbf{R}ole} is used to define the binding restrictions with the relevant \textit{Relationship}. One \textit{Relationship} is associated with two \textit{Roles}. Through each role, the \textit{Relationship} is defined as one that binds with one \textit{Point} or one \textit{Object} in its one end. 
\item \textit{\textbf{P}roperty} is a specific attribute of meta-models that is attached to the other five meta-meta models.
\item \textit{\textbf{E}xtension} refers to the additional \textit{constraints} used to construct meta-models. In this paper, one constraint is developed as a \textit{connector}. It refers to one binding between one \textit{Point} or \textit{Object} and one \textit{Role} in one side of the \textit{Relationship}.
\end{itemize}

\begin{table} 
\newcommand{\tabincell}[2]{\begin{tabular}{@{}#1@{}}#2\end{tabular}}
\centering
\caption{Interrelationships among the GOPPRR elements}
\label{TObP}
\resizebox{0.5\textwidth}{!}{ 
\begin{tabular}{cccccc}
\toprule
GOPPRR	& $\mathit{graph^{id}_a}$ & $\mathit{object^{id}_b}$ 	& $\mathit{relationship^{id}_d}$	& $\mathit{role^{id}_c}$ 	& $\mathit{point^{id}_e}$ \\\midrule
$\mathit{graph^{id}_a}$& - &\tabincell{c}{decompose\\explore}& \tabincell{c}{decompose\\explore} &explore& \tabincell{c}{decompose\\explore}\\\midrule
$\mathit{object^{id}_b}$&include&-&-&connect&-\\\midrule
$\mathit{relationship^{id}_d}$&include&-&-&-&-\\\midrule
$\mathit{role^{id}_c}$&-&-&\tabincell{c}{startFrom\\endTo}&-&-\\\midrule
$\mathit{point^{id}_e}$&-&include&- &connect&-\\\midrule
$\mathit{property^{id}_f}$& have& have& have& have& have\\\bottomrule
\end{tabular}
}
\end{table}

\newtheorem{definition}{Definition}
\begin{definition}
Token $::=$ refers to a collection of elements.
\end{definition} 
As shown in Table \ref{TObP}, the GOPPRRE meta-meta models are identified, and their interrelationships are defined. Thus, the meta-model, $Graph$, is defined as

\begin{equation} \small
\begin{aligned}
{graph}_{Tp}::=\{\sum \textit{object}^{obTp}, \sum \textit{relationship}^{reTp}, \\ \sum \textit{role}_{reTp}^{roTp}, \sum \textit{point}_{obTp}^{poTp}, \sum \textit{property}_{nonPro}^{proTp})\}
\end{aligned},
\end{equation}
where ${graph}_{Tp}$ refers to the ontological concept of a meta-model, $Graph$, whose type is defined as $Tp$. $\textit{object}_{obTp}$ refers to the ontological concept of the meta-model, $object$, where $obTp$ is a type of $object$. The $relationship^{reTp}$ refers to the ontological concept of meta-model $relationship$, where $reTp$ is a type of $Relationship$. $role_{reTp}^{roTp}$ refers to the ontology concept of a meta-model, $Role$, and $reTp$ 
refers to the $relationship$ that starts from (ends at) the $Role$, whose type is $roTp$. $point_{obTp}^{poTp}$ refers to the ontological concept of the meta-model, $Point$, and $obTp$ refers to the $object$, including the $point$, whose type is $poTp$. $property_{nonPro}^{proTp}$ refers to ontological concept of the meta-model, $Property$, and $nonPro$ refers to the $nonproperty$ elements ($nonproperty \subseteq \{Graph, Object, Relationship, Role, and Point\}$), having the $Property$ of type, $proTp$.

To define the connection rules among meta-models $Objects$ and $Points$ in each $Graph$, an additional constraint is defined as a $connector$:
\begin{equation} \footnotesize
\begin{aligned} 
connector(conId)::=\{\textit{relationship}^{reTp}, \textit{role}_{reTp}^{roTp}, \\ \textit{object}^{obTp} (\vee\textit{point}_{obTp}^{poTp}) \}
\end{aligned},
\end{equation}
where the $connector(conId)$ defines a rule that allows $reTp$, $roTp$, or $obTp$ (or $poTp$ in $obTp$) to be connected. 

\begin{equation} \footnotesize
\begin{aligned} 
graph_{Tp}(gId)::=\{\sum \textit{object}^{obTp}(obId),\sum \textit{relationship}^{reTp}(reId),  \\\sum \textit{Role}_{reTp}^{roTp}(reId,roId),  \sum \textit{Point}_{obTp}^{poTp}(obId,poId), \\ \sum \textit{Property}_{nonPro}^{proTp}(nonproId,proId)\}
\end{aligned}.
\end{equation}

\begin{definition}
 $graph_{Tp}(gId)$ refers to the model, \textit{gId}, based on the meta-model of \textit{Graph} \textit{Tp}. In $\mathit{graph_{Tp}(gId)}$, $object^{obTp}(obId)$ refers to the \textit{Object} instance, \textit{obId}, based on the meta-model of \textit{Object} \textit{obTp}. $relationship^{reTp}(reId)$ refers to the \textit{Relationship} instance, \textit{reId}, based on the meta-model of \textit{Relationship} \textit{reTp}. $\textit{Role}_{reTp}^{roTp}(reId,roId)$ refers to the \textit{Role} instance, \textit{roId}, based on the meta-model of \textit{Role} \textit{roTp} in the \textit{Relationship}, \textit{reId}, whose meta-model is \textit{Relationship} \textit{roTp}. $Point_{obTp}^{poTp}(obId,poId)$ refers to the \textit{Point} instance, \textit{poId}, based on the meta-model of \textit{Point} \textit{poTp} in the individual \textit{Object} \textit{obId}, whose meta-model is \textit{obTp}.
 $\textit{Property}_{nonPro}^{proTp}(nonProId, proId)$ refers to the property instance, \textit{proId}, based on the meta-model of \textit{Property} \textit{proTp} in the $nonproperty$ element, \textit{nonProId}, whose meta-model is \textit{nonPro};
\end{definition}

\begin{definition}
With the definition of $connector$, the concept $connection$ is defined as a link between $Objects$ or $Points$ in a $Graph$ model, which is realized as a $Relationship$. Token $a=>b$ is defined as a $connection$ that is linked from $a$ to $b$, created based on two $connector$ constraints. Thus, the $connection$, \textit{reTp}, refers to one link realized by the $Relationship$ individual, \textit{reTp}, in the MBSE models, which is defined as follows:
\end{definition}

\begin{equation} \footnotesize 
\begin{aligned} 
connection_{reTp}(reId)::=connector(conId')=>connector(conId)\\ = \{\textit{object}^{obTp'}(obId')(\vee  \textit{Point}_{obTp'}^{poTp'}(obId',poId')),\\  \textit{Role}_{reTp'}^{roTp'}(reId',roId'), \textit{relationship}^{reTp}(reId)\}\\=>\{\textit{relationship}^{reTp}(reId),\\ \textit{Role}_{reTp}^{roTp}(reId,roId), \textit{object}^{obTp}(obId)(\vee \textit{Point}_{obTp}^{poTp}(obId,poId))\}
\end{aligned},
\end{equation}
where the $connection$ is defined for $Relationship$ instance \textit{reId} whose type is $reTp$ based on $connector(conId')$ and $connector(conId)$.

\subsection{GOPPRRE Concept Mappings to Knowledge Graph Models}

\begin{figure}[!t]
\centering
\includegraphics[width=3.4in]{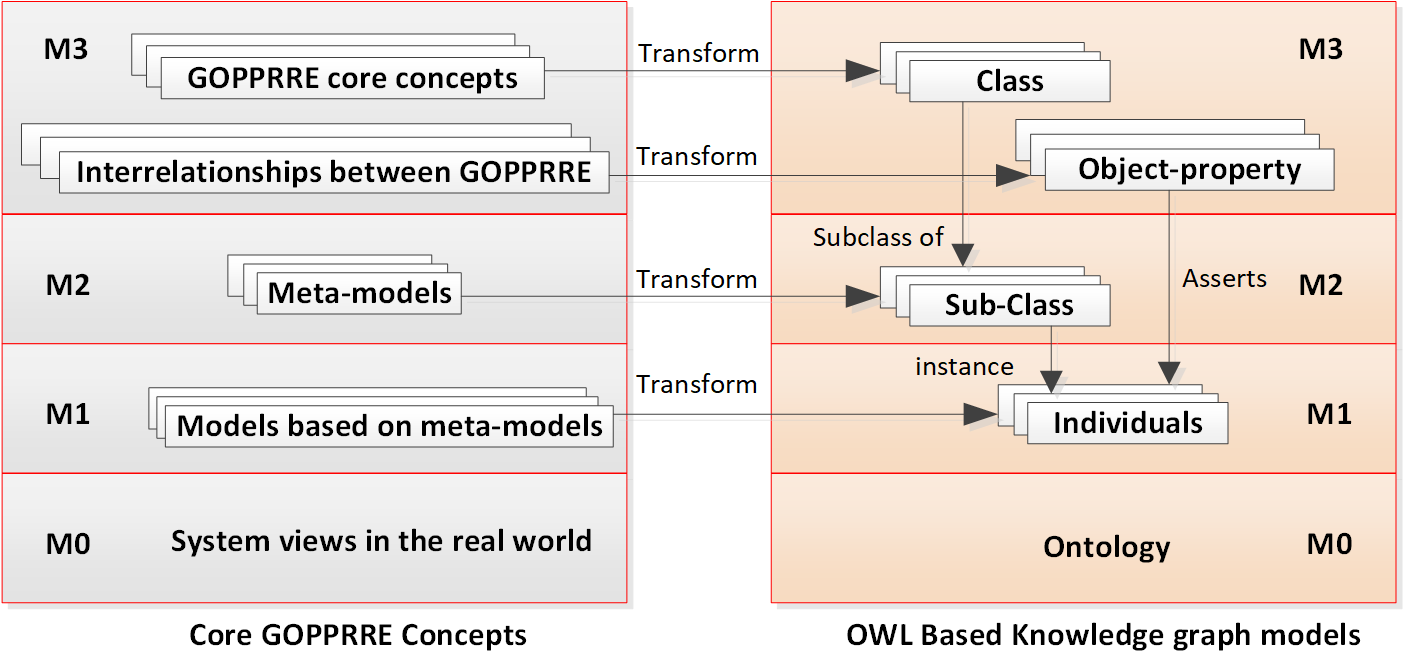}
\caption{Abstract syntax transformations}
\label{F:AST}
\end{figure}

As shown in Fig. \ref{F:AST}, a workflow for transforming the GOPPRRE core concepts to knowledge-graph models based on OWL is demonstrated. The class for each GOPPRR concept represents the GOPPRRE meta-meta models (i.e., Graph, Object, Relationship, Role, Property, Point, and Connector). Their interrelationships are transformed to object-property concepts in the knowledge graph model. Meta-models based on each GOPPRRE concept are then transformed to sub-class concepts. Models are transformed to individuals based on their related sub-classes. Based on the object-property concepts, the interrelationships among individuals are defined. Moreover, the data property is used to define the value of each $Property$. The data property type is used to define the data type of each attribute. Finally, the MBSE models representing the real-world views are transformed to the ontology defined by the knowledge-graph models using individuals, data properties, and object properties.

Apart from the abstract syntax, the concrete syntax of meta-models and models is described by the annotation and data properties.
\begin{itemize}
\item \textit{annotation property}: used to represent the abstract syntax of meta-models, such as their original \textit{icon paths}.
\item \textit{data property}: used to define the abstract syntax of models, such as the \textit{icon path} of objects in the models. This differs from the annotation property, because, when building MBSE models, the original icon of meta-models may be reconfigured. 
\end{itemize}

\begin{figure*}[h]
\centering
\includegraphics[width=6.8in]{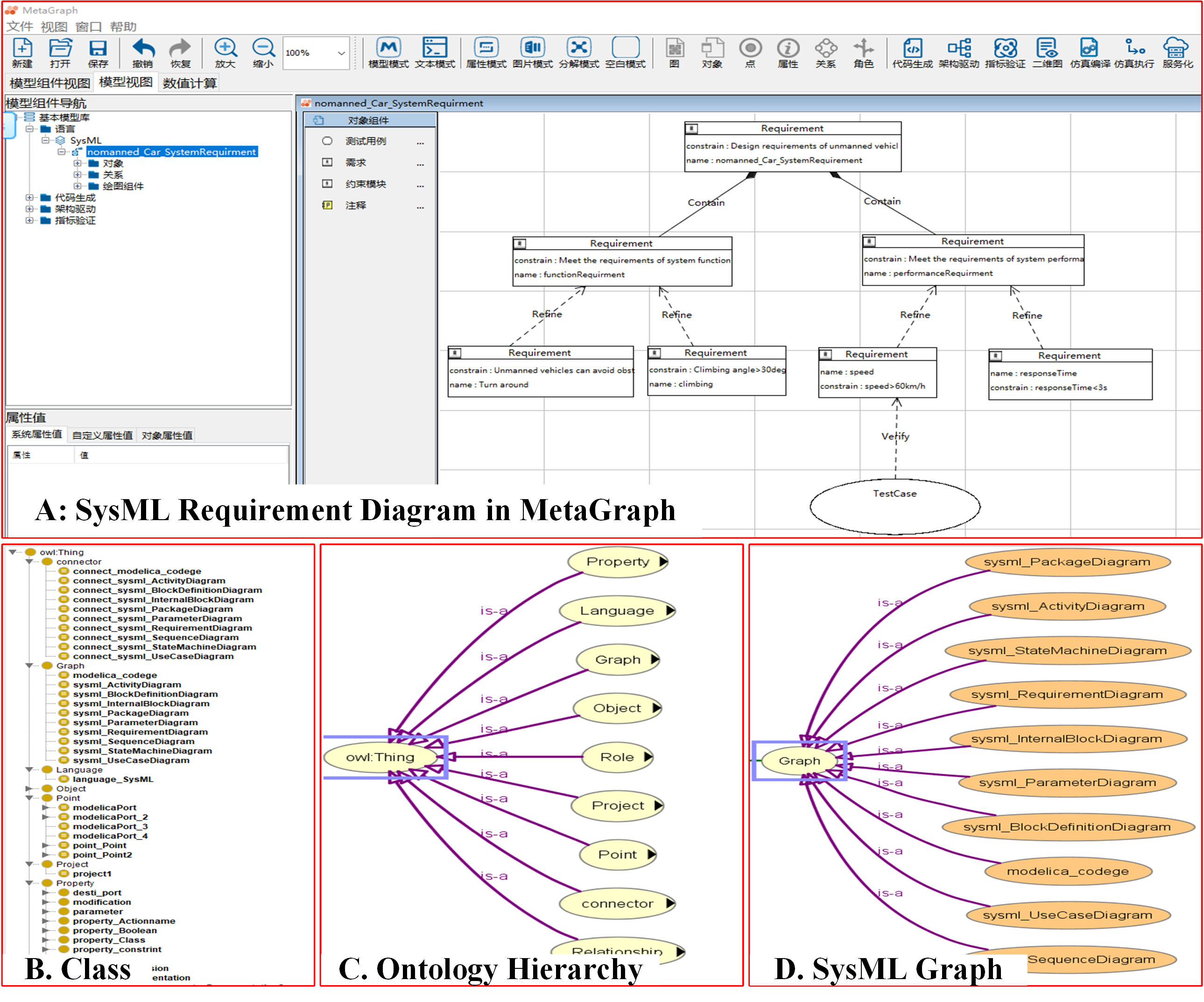}
\caption{Ontology modeling using MetaGraph}
\label{F:ONEXAM}
\end{figure*}

\section{Case Study}

Quantitative and qualitative analyses were performed to evaluate the completeness of the concrete syntax and logic of the abstract syntax. During quantitative analysis, a domain-specific modeling tool, \textit{MetaGraph}, was developed to evaluate the ontology using several MBSE languages \cite{LU202020202020}, as shown in Fig. \ref{F:ONEXAM}. Several meta-models were developed with the \textit{MetaGraph} based on five existing MBSE language specifications. In the qualitative approach, SQWRL and SPARQL were used to evaluate the completeness and logic of the developed ontology through reasonings. 

\subsection{Quantitative analysis}

When implementing the quantitative analysis, the \textit{MetaGraph} was used to develop MBSE models based on the proposed ontology, as shown in Fig. \ref{F:ONEXAM}. Moreover, five general MBSE languages were developed to evaluate whether the ontology could provide enough information for the MBSE constructions. As shown in Table \ref{TQU1}, meta-models of five general MBSE languages were built to compare the four existing tools. 

\begin{table*}[h]
\centering
\caption{Meta models described by the GOPPRRE approach.}
\label{TQU1}
\newcommand{\tabincell}[2]{\begin{tabular}{@{}#1@{}}#2\end{tabular}}
\resizebox{0.9\textwidth}{!}{
\begin{tabular}{llllllll}\hline
Language & Graph & Object & Point & Property & Relationship & Role & Referred tools \\\hline
SysML & 9(9) &73(64) & 11(11)& 10(10)& 31(31) &  31(31)& (Magic draw) \cite{Chami2018}\\\hline
BPMN & 1(1)   & 81(77)  & 0(0)  & 51(46)  & 5(3)  & 10(0)  & (BPM Camunda) \cite{Fernandez} \\\hline
\tabincell{c}{UPDM (Unified Profile\\ for DoDAF/MODAF)} & 52(52)  & 100 (123) & 7(6) & 84 (96) & 57 (50)& 54 (0)& (Magic draw)  \cite{Chami2018}\\\hline
EASTADL & 10(9) & 67(62) & 17(17) & 93(93) & 23(21) &68(64)  & (MetaEdit+)  \cite{Kelly1996}\\\hline
OPM & 1(1) & 3(3) & 0(0) & 9(8) &  15(15)&  12(4)& (OPCAT) \cite{Dori2016} \\\hline
\end{tabular}
}
\end{table*}

From the results, we found that the ontology could formalize almost all meta-models of the related languages. All graphs were developed based on the five MBSE language specifications. Some objects were different from the existing tools, because some elements in their tools were not defined as objects in our approach. For example, in Magic draw, some properties were defined as elements in their diagram-building environment so that the users could easily configure an object’s property. The concrete syntax of all languages were completely transformed to the developed ontology in \textit{MetaGraph}.

Apart from the concrete syntax, the abstract syntax was also evaluated by comparing the connection rules with different languages in other tools. In the \textit{MetaGraph}, the connectors between relationships and objects were compared with the rules for connecting different elements in other tools. This was done to determine whether the ontology can formalize the logic flow between different MBSE model elements. As shown in Table \ref{TQU2}, connection rules refer to the specifications used to define how to connect model compositions and their ports for the five existing languages in other tools. The connectors were used to create connections between \textit{Objects} and \textit{Points} in our approach. From the results, we found that almost all connection rules were defined based on the given ontology, although the number of connectors was not twice the connection rules of different MBSE languages. This occurred because of the discrepancies of constructing the \textit{Graph} meta model. For example, in BPMN, the number of connectors was 11 fewer than twice the number of connections, because the linkings between the text \textit{Object} and other 12 \textit{Objects} in the BPMN specification required one \textit{Role} for the text \textit{Object} and 12 roles for other \textit{Objects} in our approach, compared with the 12 connection rules in other tools. 

\begin{table}[h]\footnotesize
\centering
\caption{Connection rules compared with connectors in the GOPPRRE approach}
\label{TQU2}
\resizebox{0.4\textwidth}{!}{
\begin{tabular}{lll}\hline
Language & Connection rules & Connectors\\\hline
SysML & 31 & 62\\\hline
BPMN & 72 & 133\\\hline
UPDM & 396 & 792\\\hline
EASTADL & 215 & 233\\\hline
OPM & 39 & 54 \\\hline 
\end{tabular}
}
\end{table}

\subsection{Qualitative analysis}

To qualitatively verify the ontology, SPARQL and SQWRL were used to evaluate the completeness and logic of MBSE models through reasoning. As shown in Fig. \ref{F:ONEXAM}-A, a SysML model was transformed to the defined ontology (Fig. \ref{F:ONEXAM}-B, C, and D) generated by \textit{MetaGraph}. The completeness and logic of the given SysML model were evaluated separately using Algorithms 1 and 2.

\begin{algorithm} 
\caption{SPARQL Algorithm for verifying the completeness of the MBSE models }
\label{alg:SQWRLLCCM}
\begin{algorithmic}
\STATE {
$\mathit{PREFIX owl:<http://www.w3.org/2002/07/owl\#> }$\newline
$\mathit{PREFIX rdf:<http://www.w3.org/1999/02/22-rdf}$\newline$\mathit{-syntax-ns\#>}$\newline
$\mathit{PREFIX xsd:<http://www.w3.org/2001/}$\newline$ \mathit{XMLSchema\#>}$\newline
$\mathit{PREFIX se\:<http://www.zkhoneycomb.com/formats/}$\newline $\mathit{metagInOwl\#>}$\newline
}

\STATE $//$ If Graph includes Objects(Relationships)
\STATE {
select ?graph ?object ?relationship\newline
where \{\newline
 ?graph se:graphIncludingObject(graphIncludingRelationship)   ?object(relationship)   \newline
\}    \newline
}
\STATE $//$ If Objects(Relationships) includes Points(Roles)
\STATE {

select ?object ?point ?relationship ?role \newline
where \{\newline
 ?object(relationship) se:linkObjectAndPoint(linkRelationship\newline AndRole) ?point(role)

\}    \newline
}
\STATE $//$ If Object(Graph and Relationship) includes Properties

\STATE {

select ?graph ?object ?point ?relationship ?role ?property\newline
where \{\newline
 ?graph(object, point, relationship or role) se:hasProperty ?Property

\}    \newline
}

\end{algorithmic}
\end{algorithm}

\begin{figure*}[h]
\centering
\includegraphics[width=6.8in]{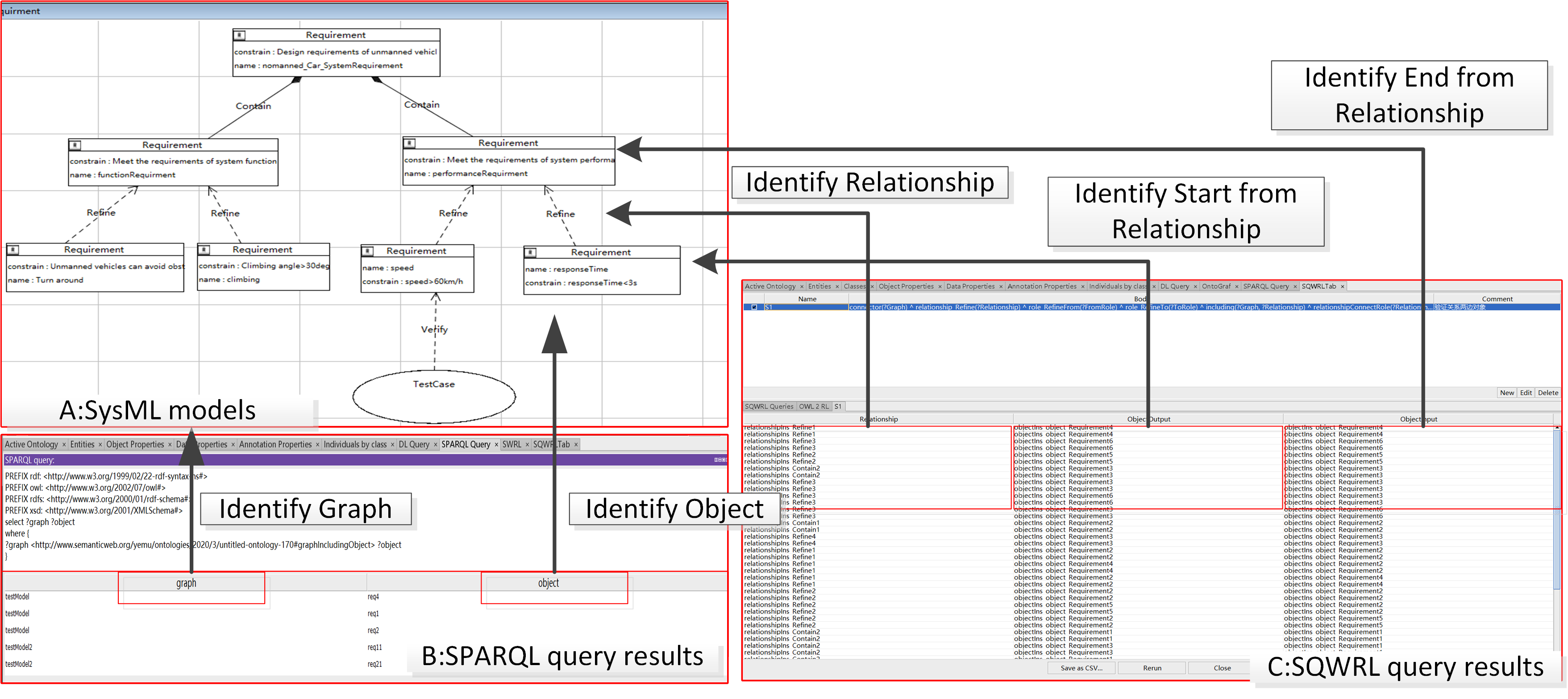}
\caption{SPARQL and SQWRL query results}
\label{F:rest123}
\end{figure*}

Algorithm \ref{alg:SQWRLLCCM} is a SPARQL query algorithm developed to verify completeness of the generated ontology. As shown in Fig. \ref{F:ONEXAM}, the ontology generated from the SysML model was used to verify the completeness of the ontology. Based on Algorithm 1, the SPARQL script was developed to verify the three metrics mentioned in Section II. As shown in Fig. \ref{F:rest123}-B, the query results demonstrated that all $Objects$ and $Relationships$ representing the SysML model were captured in the ontology to describe its model structure. Moreover, $Properties$ was also identified in different meta-models. From the results, we can infer that the completeness of the ontology was verified, because all the information related to the SysML model was completely transformed into the ontology model.

\begin{algorithm} 
\caption{SQWRL Algorithm for verifying the logic of the MBSE models}
\label{alg:lcdsm}
\begin{algorithmic}
\STATE $//$ Query relationships in the MBSE models
\STATE {graph(?Graph) $\Lambda$ connector(?Connector1) $\Lambda$ connector(?Connector2)  $\Lambda$ relationship(?Relationship)  $\Lambda$ object(?ObjectInput)  $\Lambda$ object(?ObjectOutput) $\Lambda$ graphIncludingConnector(?Graph, ?Connector1)  $\Lambda$  graphIncludingConnector(?Graph, ?Connector2) $\Lambda$  linkFromRelationship(?Connector1,?Relationship)  $\Lambda$  linkFromRelationship(?Connector2,?Relationship)  $\Lambda$   linkToObject(?Connector1,? ObjectInput) $\Lambda$ linkToObject(?Connector2, ?ObjectOutPut) $\Lambda$ connect(?Connector1, ?Connector2) 
}
\STATE {$->$ sqwrl:select(?Relationship,?ObjectInput,?ObjectOutput)}

\STATE $//$ Query the direction of each relationship 

\STATE {$->$ sqwrl:select(?Graph, ?Relationship, ?ObjectInput)}

\end{algorithmic}
\end{algorithm}

Algorithm \ref{alg:lcdsm} is a SQWRL algorithm used to verify the logic flows in the given SysML model. To capture the $connections$ among $\mathit{object^{obTp}(obId)}$, $\mathit{relationship^{reTp}(obId)}$, and $\mathit{point^{reTp}(obId,poId)}$, which are defined as the individuals representing the the \textit{Object}, \textit{Relationship}, and \textit{Point} concepts in the model, Algorithm \ref{alg:lcdsm} was used to capture the related information. All individuals representing the SysML model were queried using the object properties listed below:
\begin{itemize}
\item \textit{graphIncludingConnector} refers to the connector developed in a graph associated with \textit{Relationship}, \textit{Role}, and \textit{Object} (\textit{Points}), as shown in Equation (2).
\item \textit{linkFromRelationship} refers to the relationship linked to one connector, where one \textit{Relationship} has one \textit{Role} as its end, as described by \textit{linkRelationshipAndRole}.
\item \textit{linkToObject} refers to the connector linked to one \textit{Object}, where one \textit{Role} is connected to one \textit{Object} or one \textit{Point} described by \textit{roleBindingObject} or \textit{roleBindingPoint}. If \textit{Points} are not involved in the connection, the \textit{Object} is defined as the end of the relationship or vice versa. 
\item \textit{connect} refers to that one connector (start) linked to another connector (end). It is used to describe the direction of the relationship.
\end{itemize} 

As shown in Fig. \ref{F:rest123}-C, the query results identify the \textit{Relationship} individuals between different \textit{Object} individuals and \textit{Point} individuals. Moreover, the direction of the \textit{Relationship} is identified based on its starting role from \textit{Object} individuals.

\subsection{Discussion}

From the quantitative and qualitative analyses, we found that the ontology based on the GOPPRRE approach could formalize at least five MBSE modeling languages used to model systems of systems (e.g., UPDM), system architectures (e.g., SysML), business processes (e.g., BPMN), and domain-specific knowledge for the architectural description language of automotive embedded systems, for example. Thus, we can infer that the designed ontology can support the MBSE formalisms for the entire lifecycle.

This ontology enables the promotion of data interoperability. \textit{GOPPRR} provides one the most powerful approaches available to describe domain-specific characteristics, whose meta-meta models have better descriptive capabilities \cite{Kern2011} than others. Moreover, from the results shown in Tables \ref{TQU1} and \ref{TQU2}, we found that the current GOPPRRE ontology could integrate at least five existing MBSE languages. To support data exchange among these languages, the GOPPRRE ontology can be used as the middleware 
for the MBSE community.

OWL is widely used to support ML and AI techniques, and its ontology, generated from MBSE models, can be used to support reasoning and to analyze target modeling systems. For example, Algorithms I and II enable information capture from MBSE models for knowledge management. Moreover, the ontology generated from MBSE models are directly used to construct cognitive twins to support decision-making, system development, and operations \cite{JINZHIct}. With this ontology, AI and ML algorithms can be developed to support system development based on MBSE model information.

Scalability is an important issue regarding the application of the developed ontology. Thus, it will eventually require adoption by the Industrial Ontologies Foundry Systems Engineering Working Group.

\section{Conclusion}

In this paper, we designed an ontology based on the GOPPRRE approach that supports MBSE formalisms using OWL methodologies for model integration. First, we demonstrated the GOPPRRE concepts using an M0--M3 modeling framework. Then, we developed a transformation rule between them and ontology based on OWL. Based on the transformation rules, OWL models were generated from five existing MBSE languages by a domain-specific modeling tool \textit{MetaGraph}. Qualitative and quantitative approaches were used to evaluate the completeness and logic of the generated ontology models. From the results, we found that the designed ontology could support MBSE formalisms, showing the potential of this method to become the standardized ontology for the MBSE community in the future.

\section*{Acknowledgment}

The work presented in this paper was supported by the EU H2020 project (869951) FACTLOG-Energy-aware Factory Analytics for Process Industries and EU H2020 project (825030) QU4LITY Digital Reality in Zero Defect Manufacturing and the InnoSwiss IMPULSE project on Digital Twins.

\ifCLASSOPTIONcaptionsoff
  \newpage
\fi

\bibliographystyle{IEEEtran}
\bibliography{MODELSBIB} 

\begin{thebibliography}{10}
\providecommand{\url}[1]{#1}
\csname url@samestyle\endcsname
\providecommand{\newblock}{\relax}
\providecommand{\bibinfo}[2]{#2}
\providecommand{\BIBentrySTDinterwordspacing}{\spaceskip=0pt\relax}
\providecommand{\BIBentryALTinterwordstretchfactor}{4}
\providecommand{\BIBentryALTinterwordspacing}{\spaceskip=\fontdimen2\font plus
\BIBentryALTinterwordstretchfactor\fontdimen3\font minus
  \fontdimen4\font\relax}
\providecommand{\BIBforeignlanguage}[2]{{%
\expandafter\ifx\csname l@#1\endcsname\relax
\typeout{** WARNING: IEEEtran.bst: No hyphenation pattern has been}%
\typeout{** loaded for the language `#1'. Using the pattern for}%
\typeout{** the default language instead.}%
\else
\language=\csname l@#1\endcsname
\fi
#2}}
\providecommand{\BIBdecl}{\relax}
\BIBdecl

\bibitem{9062568}
M.~{Kharrat}, O.~{Penas}, R.~{Plateaux}, J.~{Choley}, H.~{Trabelsi},
  J.~{Louati}, and M.~{Haddar}, ``Integration of electromagnetic constraints as
  of the conceptual design through an mbse approach,'' \emph{IEEE Systems
  Journal}, pp. 1--12, 2020.

\bibitem{McDermott2020}
T.~McDermott, D.~DeLaurentis, P.~Beling, M.~Blackburn, and M.~Bone, ``{AI4SE
  and SE4AI: A Research Roadmap},'' \emph{Insight}, vol.~23, no.~1, pp. 8--14,
  2020.

\bibitem{Hao2019}
\BIBentryALTinterwordspacing
J.~Hao, L.~Xu, G.~Wang, Y.~Jin, and Y.~Yan, ``{A Knowledge-Based Method for
  Rapid Design Concept Evaluation},'' \emph{IEEE Access}, vol.~7, pp.
  116\,835--116\,847, 2019. [Online]. Available:
  \url{https://ieeexplore.ieee.org/document/8790704/}
\BIBentrySTDinterwordspacing

\bibitem{Ming2018}
\BIBentryALTinterwordspacing
Z.~Ming, G.~Wang, Y.~Yan, J.~H. Panchal, C.~H. Goh, J.~K. Allen, and
  F.~Mistree, ``{Ontology-Based Representation of Design Decision
  Hierarchies},'' \emph{Journal of Computing and Information Science in
  Engineering}, vol.~18, no.~1, mar 2018. [Online]. Available:
  \url{https://asmedigitalcollection.asme.org/computingengineering/article/doi/10.1115/1.4037934/366507/OntologyBased-Representation-of-Design-Decision}
\BIBentrySTDinterwordspacing

\bibitem{Wang2019}
H.~Wang, G.~Wang, J.~Lu, and C.~Ma, ``{Ontology Supporting Model-Based Systems
  Engineering Based on a GOPPRR Approach},'' in \emph{WorldCist'19 - 7th World
  Conference on Information Systems and Technologies}.\hskip 1em plus 0.5em
  minus 0.4em\relax Cham: Springer International Publishing, 2019, pp.
  426--436.

\bibitem{Dickerson2013}
\BIBentryALTinterwordspacing
C.~E. Dickerson and D.~Mavris, ``{A Brief History of Models and Model Based
  Systems Engineering and the Case for Relational Orientation},'' \emph{IEEE
  Systems Journal}, vol.~7, no.~4, pp. 581--592, dec 2013. [Online]. Available:
  \url{http://ieeexplore.ieee.org/document/6506950/}
\BIBentrySTDinterwordspacing

\bibitem{Bassi2011}
\BIBentryALTinterwordspacing
L.~Bassi, C.~Secchi, M.~Bonfe, and C.~Fantuzzi, ``{A SysML-Based Methodology
  for Manufacturing Machinery Modeling and Design},'' \emph{IEEE/ASME
  Transactions on Mechatronics}, vol.~16, no.~6, pp. 1049--1062, dec 2011.
  [Online]. Available: \url{http://ieeexplore.ieee.org/document/5604318/}
\BIBentrySTDinterwordspacing

\bibitem{Mordecai2018}
Y.~Mordecai, O.~Orhof, and D.~Dori, ``{Model-Based Interoperability Engineering
  in Systems-of-Systems and Civil Aviation},'' \emph{IEEE Transactions on
  Systems, Man, and Cybernetics: Systems}, vol.~48, no.~4, pp. 637--648, apr
  2018.

\bibitem{Ravikumar2018}
\BIBentryALTinterwordspacing
G.~Ravikumar, S.~A. Khaparde, and R.~K. Joshi, ``{Integration of Process Model
  and CIM to Represent Events and Chronology in Power System Processes},''
  \emph{IEEE Systems Journal}, vol.~12, no.~1, pp. 149--160, mar 2018.
  [Online]. Available: \url{http://ieeexplore.ieee.org/document/7448852/}
\BIBentrySTDinterwordspacing

\bibitem{Kelly1996}
\BIBentryALTinterwordspacing
S.~Kelly, K.~Lyytinen, and M.~Rossi, \emph{{Advanced Information Systems
  Engineering}}, ser. Lecture Notes in Computer Science, P.~Constantopoulos,
  J.~Mylopoulos, and Y.~Vassiliou, Eds.\hskip 1em plus 0.5em minus 0.4em\relax
  Berlin, Heidelberg: Springer Berlin Heidelberg, 1996, vol. 1080. [Online].
  Available:
  \url{http://www.springerlink.com/content/fj603m1wu231v560/?p=0c78bbba5cb14fbaa4a45c2a75e6c474{\&}pi=0
  http://link.springer.com/10.1007/3-540-61292-0}
\BIBentrySTDinterwordspacing

\bibitem{Nuzzo2015}
\BIBentryALTinterwordspacing
P.~Nuzzo, A.~L. Sangiovanni-Vincentelli, D.~Bresolin, L.~Geretti, and T.~Villa,
  ``{A Platform-Based Design Methodology With Contracts and Related Tools for
  the Design of Cyber-Physical Systems},'' \emph{Proceedings of the IEEE}, vol.
  103, no.~11, pp. 2104--2132, nov 2015. [Online]. Available:
  \url{http://ieeexplore.ieee.org/document/7268792/}
\BIBentrySTDinterwordspacing

\bibitem{Beihoff2010}
B.~Beihoff, S.~Friedenthal, D.~Kemp, D.~Nichols, C.~Oster, C.~Peredis,
  H.~Stoewer, and J.~Wade, ``{A World in Motion Systems Engineering Vision
  2025},'' in \emph{International Council on Systems Engineering}, vol. 327,
  no. 5970, mar 2010, pp. 1183--1183.

\bibitem{Yang2020}
\BIBentryALTinterwordspacing
L.~Yang, K.~Cormican, and M.~Yu, ``{Ontology Learning for Systems Engineering
  Body of Knowledge},'' \emph{IEEE Transactions on Industrial Informatics},
  vol. 3203, no.~c, pp. 1--1, 2020. [Online]. Available:
  \url{https://ieeexplore.ieee.org/document/9079664/}
\BIBentrySTDinterwordspacing

\bibitem{Seidner2008}
\BIBentryALTinterwordspacing
C.~Seidner and O.~Roux, ``{Formal Methods for Systems Engineering Behavior
  Models},'' \emph{IEEE Transactions on Industrial Informatics}, vol.~4, no.~4,
  pp. 280--291, nov 2008. [Online]. Available:
  \url{http://ieeexplore.ieee.org/document/4689308/}
\BIBentrySTDinterwordspacing

\bibitem{Lu2020ieeeesss}
\BIBentryALTinterwordspacing
J.~Lu, G.~Wang, and M.~Torngren, ``{Design Ontology in a Case Study for
  Cosimulation in a Model-Based Systems Engineering Tool-Chain},'' \emph{IEEE
  Systems Journal}, vol.~14, no.~1, pp. 1297--1308, mar 2020. [Online].
  Available: \url{https://ieeexplore.ieee.org/document/8734748/}
\BIBentrySTDinterwordspacing

\bibitem{Schluse2018}
M.~Schluse, M.~Priggemeyer, L.~Atorf, and J.~Rossmann, ``{Experimentable
  Digital Twins-Streamlining Simulation-Based Systems Engineering for Industry
  4.0},'' \emph{IEEE Transactions on Industrial Informatics}, vol.~14, no.~4,
  pp. 1722--1731, 2018.

\bibitem{Hao2014}
\BIBentryALTinterwordspacing
J.~Hao, Y.~Yan, L.~Gong, G.~Wang, and J.~Lin, ``{Knowledge map-based method for
  domain knowledge browsing},'' \emph{Decision Support Systems}, vol.~61, pp.
  106--114, may 2014. [Online]. Available:
  \url{https://linkinghub.elsevier.com/retrieve/pii/S0167923614000244}
\BIBentrySTDinterwordspacing

\bibitem{JINZHIct}
J.~Lu, X.~Zheng, A.~Gharaei, K.~Kalaboukas, and D.~Kiritsis, ``Cognitive twins
  for supporting decision-makings of internet of things systems,'' in
  \emph{Proceedings of 5th International Conference on the Industry 4.0 Model
  for Advanced Manufacturing}, L.~Wang, V.~D. Majstorovic, D.~Mourtzis,
  E.~Carpanzano, G.~Moroni, and L.~M. Galantucci, Eds.\hskip 1em plus 0.5em
  minus 0.4em\relax Cham: Springer International Publishing, 2020, pp.
  105--115.

\bibitem{Papakonstantinou2013}
\BIBentryALTinterwordspacing
N.~Papakonstantinou and S.~Sierla, ``{Generating an Object Oriented IEC 61131-3
  software product line architecture from SysML},'' in \emph{2013 IEEE 18th
  Conference on Emerging Technologies {\&} Factory Automation (ETFA)}.\hskip
  1em plus 0.5em minus 0.4em\relax IEEE, sep 2013, pp. 1--8. [Online].
  Available: \url{http://ieeexplore.ieee.org/document/6648057/}
\BIBentrySTDinterwordspacing

\bibitem{Walter2014}
\BIBentryALTinterwordspacing
T.~Walter, F.~S. Parreiras, and S.~Staab, ``{An ontology-based framework for
  domain-specific modeling},'' \emph{Software {\&} Systems Modeling}, vol.~13,
  no.~1, pp. 83--108, feb 2014. [Online]. Available:
  \url{http://link.springer.com/10.1007/s10270-012-0249-9}
\BIBentrySTDinterwordspacing

\bibitem{Patti2016}
\BIBentryALTinterwordspacing
E.~Patti, A.~Acquaviva, M.~Jahn, F.~Pramudianto, R.~Tomasi, D.~Rabourdin,
  J.~Virgone, and E.~Macii, ``{Event-Driven User-Centric Middleware for
  Energy-Efficient Buildings and Public Spaces},'' \emph{IEEE Systems Journal},
  vol.~10, no.~3, pp. 1137--1146, sep 2016. [Online]. Available:
  \url{http://ieeexplore.ieee.org/lpdocs/epic03/wrapper.htm?arnumber=6740031}
\BIBentrySTDinterwordspacing

\bibitem{LU202020202020}
L.~Jinzhi, W.~Guoxin, M.~Junda, K.~Dimitris, Z.~Hang, and T.~Martin, ``{General
  Modeling Language to Support Model-based Systems Engineering Formalisms (Part
  1)},'' in \emph{INCOSE International Symposium 30 (In press)}, 2020.

\bibitem{OConnor2009}
M.~O'Connor and A.~Das, ``{SQWRL: A query language for OWL},'' in \emph{CEUR
  Workshop Proceedings}, 2009.

\bibitem{horrocks2004swrl}
I.~Horrocks, P.~F. Patel-Schneider, H.~Boley, S.~Tabet, B.~Grosof, M.~Dean
  \emph{et~al.}, ``{SWRL: A Semantic Web Rule Language Combining OWL and
  RuleML},'' \emph{W3C Member submission}, vol.~21, p.~79, 2004.

\bibitem{Lu2018aaaa}
\BIBentryALTinterwordspacing
J.~Lu, D.~G{\"{u}}rd{\"{u}}r, D.-J. Chen, J.~Wang, and M.~T{\"{o}}rngren,
  ``{Empirical-Evolution of Frameworks Supporting Co-simulation Tool-Chain
  Development},'' in \emph{Advances in Intelligent Systems and Computing},
  2018, vol. 745, pp. 813--828. [Online]. Available:
  \url{http://link.springer.com/10.1007/978-3-319-77703-0{\_}80}
\BIBentrySTDinterwordspacing

\bibitem{Mierlo2017}
\BIBentryALTinterwordspacing
S.~V. Mierlo, Y.~V. Tendeloo, B.~Meyers, and H.~Vangheluwe,
  \emph{{Domain-Specific Modelling for Human–Computer Interaction}}, ser.
  Human–Computer Interaction Series, B.~Weyers, J.~Bowen, A.~Dix, and
  P.~Palanque, Eds.\hskip 1em plus 0.5em minus 0.4em\relax Cham: Springer
  International Publishing, 2017. [Online]. Available:
  \url{http://link.springer.com/10.1007/978-3-319-51838-1}
\BIBentrySTDinterwordspacing

\bibitem{Chami2018}
\BIBentryALTinterwordspacing
M.~Chami, A.~Aleksandraviciene, A.~Morkevicius, and J.-m. Bruel, ``{Towards
  Solving MBSE Adoption Challenges: The D3 MBSE Adoption Toolbox},''
  \emph{INCOSE International Symposium}, vol.~28, no.~1, pp. 1463--1477, jul
  2018. [Online]. Available:
  \url{http://doi.wiley.com/10.1002/j.2334-5837.2018.00561.x}
\BIBentrySTDinterwordspacing

\bibitem{Fernandez}
A.~Fernandez and A.~Fernandez, ``{Camunda BPM Platform Loan Assessment Process
  Lab},'' Tech. Rep.

\bibitem{Dori2016}
\BIBentryALTinterwordspacing
D.~Dori, \emph{{Model-Based Systems Engineering with OPM and SysML}}.\hskip 1em
  plus 0.5em minus 0.4em\relax New York, NY: Springer New York, 2016. [Online].
  Available: \url{http://link.springer.com/10.1007/978-1-4939-3295-5}
\BIBentrySTDinterwordspacing

\bibitem{Kern2011}
H.~Kern, A.~Hummel, and S.~K{\"{u}}hne, ``{Towards a comparative analysis of
  meta-metamodels},'' in \emph{Proceedings of the compilation of the co-located
  workshops on DSM11}, vol.~1, 2011, pp. 7--12.

\end{thebibliography}

\begin{IEEEbiography}[{\includegraphics[width=1in,height=1.25in,clip,keepaspectratio]{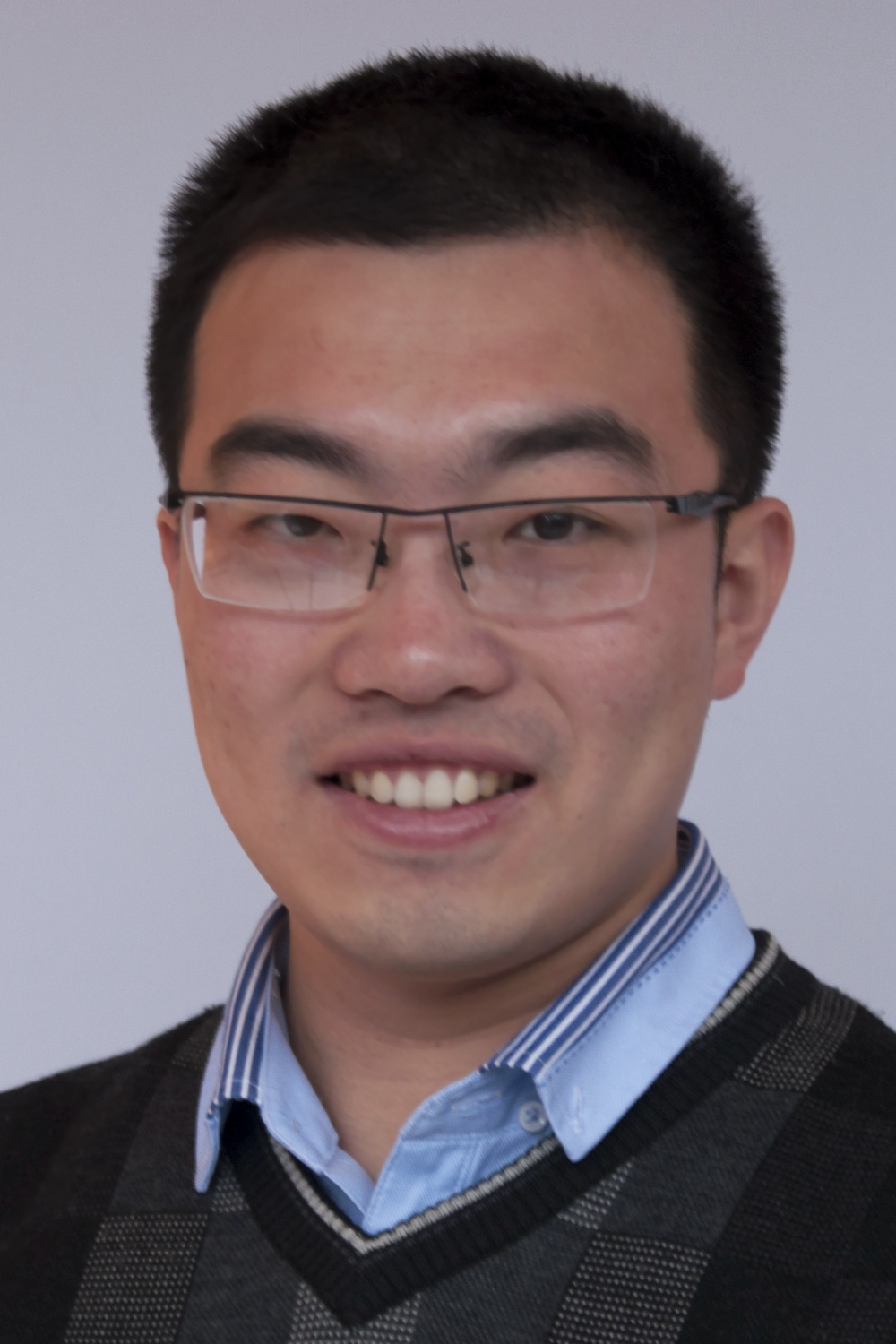}}]{Jinzhi Lu}
, CSEP, is a research scientist at EPFL. He got his ph.d degree at KTH Royal Institute of Technology, Mechatronics Division in 2019. His research interest is MBSE tool-chain design and MBSE enterprise transitioning. He is senior member of China Council on Systems Engineering (CCOSE), China Council on Systems Engineering.
\end{IEEEbiography}

\begin{IEEEbiography}[{\includegraphics[width=1in,height=1.25in,clip,keepaspectratio]{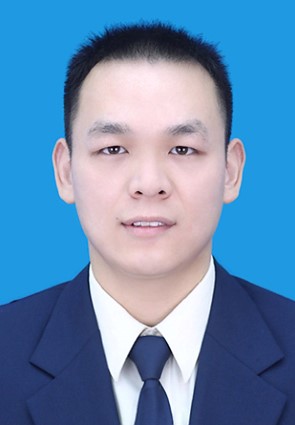}}]{Junda Ma}
 is a Ph.D.student at the Institute of Industrial and Intelligent System Engineering, School of Mechanical Engineering, Beijing Institute of Technology, majoring in mechanical engineering. His research interests are MBSE tool design and R\&D of multi-architecture modeling language.
\end{IEEEbiography}

\begin{IEEEbiography}[{\includegraphics[width=1in,height=1.25in,clip,keepaspectratio]{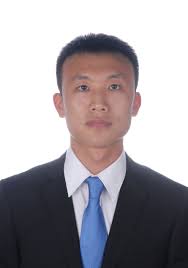}}]
{Xiaochen Zheng}
Ph.D, received his doctoral degree from Universidad Politécnica de Madrid. Before that he studied in Shandong University in Mechanical Engineering and obtained his bachelor and master degree. He is now working at École Polytechnique Fédérale de Lausanne as  a postdoctoral scientist. His research interests include Internet of Things, Machine learning, Wearable technology, Distributed ledger technology and their applications in industry and healthcare etc.
\end{IEEEbiography}

\begin{IEEEbiography}[{\includegraphics[width=1in,height=1.25in,clip,keepaspectratio]{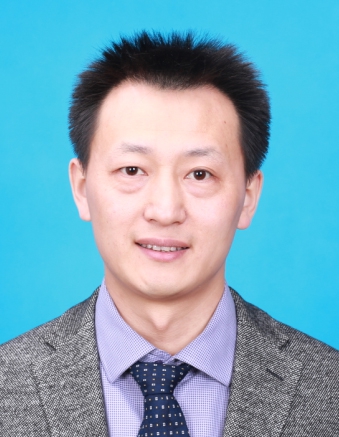}}]
{Guoxin Wang} is an associate professor and director of Industrial Engineering Institute at Beijing Institute of Technology. His research interests include Reconfigurable Manufacturing System and Knowledge based Engineering.
\end{IEEEbiography}

\begin{IEEEbiography}[{\includegraphics[width=1in,height=1.25in,clip,keepaspectratio]{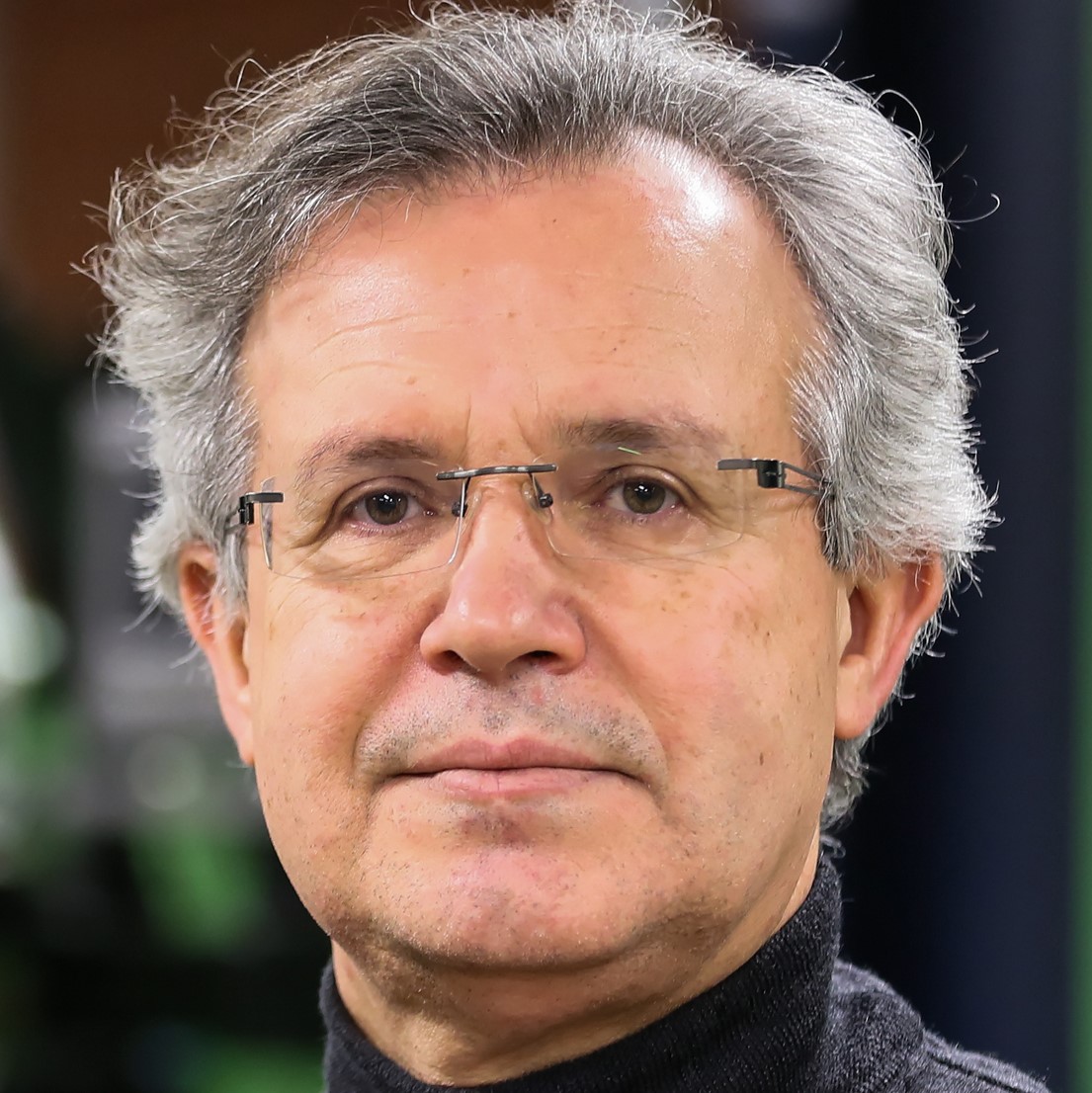}}]
{Dimitris Kiritsis} is Faculty Member at the Institute of Mechanical Engineering of the School of Engineering of EPFL, Switzerland, where he is leading a research group on ICT for Sustainable Manufacturing. He serves also as Director of the doctoral Program of EPFL on Robotics, Control and Intelligent Systems (EDRS). His research interests are Closed Loop Lifecycle Management, IoT, Semantic Technologies and Data Analytics for Engineering Applications. 
\end{IEEEbiography}

\end{document}